\documentclass[12pt]{iopart}

\usepackage{iopams}  

\usepackage{theorem}
\usepackage[dvipdfmx]{graphicx}

\theoremstyle{break}
\newtheorem{Definition}{Definition}

\newtheorem{Theorem}{Theorem}
\newtheorem{Proof}{Proof}

\newtheorem{Proposition}[Theorem]{Proposition}
\newtheorem{Corollary}[Theorem]{Corollary}
\newtheorem{Lemma}[Theorem]{Lemma}
\newtheorem{Example}{Example}
\newtheorem{Fact}[Theorem]{Fact}

\def\ds{\displaystyle}
\def\qed{\hfill\hbox{$\Box$}\vspace{0pt}\break}

\def\F{{\mathbb F}}
\def\R{{\mathbb R}}
\def\Z{{\mathbb Z}}

\def\1{\boldsymbol{1}}

\def\calL{{\cal L}}

\def\om{\omega}
\def\Om{\Omega}

\begin{document}

\title{Solitons with nested structure over finite fields}

\author{Fumitaka Yura}

\address{
  Department of Complex and Intelligent Systems,
  Future University HAKODATE, 
  116-2 Kamedanakano-cho Hakodate Hokkaido, 041-8655, Japan
}
\ead{yura@fun.ac.jp}
\begin{abstract}
We propose a solitonic dynamical system over finite fields
  that may be regarded as an analogue of the box-ball systems.
The one-soliton solutions of the system, which have nested structures similar to fractals, 
  are also proved. 
The solitonic system in this paper is described by polynomials, which seems to be novel.
Furthermore, in spite of such complex internal structures,
  numerical simulations exhibit stable propagations before and after collisions
  among multiple solitons with preserving their patterns.
\end{abstract}

%\maketitle

\section{Introduction}
\label{sec:intro}

The box-ball system (BBS)\cite{TS90} has been extensively studied as {\it digitalized} integrable systems. 
The properties of the BBS such as the relation to the discrete KP (or KdV) equation
  and its conserved quantities etc.~has been well understood through the ultradiscretization\cite{TTMS}. 
From the ultradiscretized system, we may frequently obtain systems such as cellular automata over integers
  by limiting the parameters and initial values to integers. 
This limitation is, however, not necessarily required for the equation itself. 
The ultradiscretized equations may be essentially regarded as the systems over
  real or rational numbers\cite{Hirota97}. 
In many cases of the integrable ultradiscrete systems that has been hitherto known, 
  their integrability has a close relationship 
  with that of the corresponding discrete system before ultradiscretization; 
  therefore, the integrability of the ultradiscrete system may be considered
  to be guaranteed by that of the corresponding discrete system. 
This property is one of remarkable merits of ultradiscretization for proving the integrability
  of the given system as a matter of course. 
However, on the other hand, intrinsic qualities of ultradiscretized system seems to be
  hidden behind the outstanding ultradiscretized correspondence. 
In this paper, we consider a solitonic dynamical system in which the dependent variables are over finite fields.
By investigating integrable dynamical systems over finite fields without such an a priori relationship, 
  novel insights about integrability itself may be expected.

Below are the examples of integrable systems over finite fields. 
In \cite{Bruschi06}, filter type cellular automata (CA) are constructed from the Schr\"odinger 
discrete spectral problem\cite{Bruschi92} that have infinitely many conserved quantities.
In \cite{Bobenko93}, integrable CA over cyclic group is obtained
  from the discrete sine-Gordon equation (Hirota equation) as M\"{o}bius transformation.
These two systems possess the Lax representation.
The algebro-geometric method for constructing solutions of the discrete KP equation 
  and the discrete Toda equation over finite fields is shown in \cite{Bialecki03,Doliwa03},
  whose $\tau$ function in the bilinear form has $N$-soliton solutions.
Three examples above do not seem to have solitonically propagating waves.
In \cite{Kanki12},
  the discrete integrable system over the rational function field with indeterminate over $\F_q$
  is obtained from the generalized discrete KdV equation.
The $p$-adic valuations of variables for the discrete KdV equation
  are also discussed in \cite{Kanki13}, which is considered as an analogue of ultradiscretization.
As an application of integrable systems over finite fields, 
  the relationship between the dynamics of Toda molecule over finite fields and the BCH-Goppa decoding
  is discussed in \cite{NM98}.

In this paper, we construct a formal analogue of the BBS over finite fields
  from the point of view of ultradiscrete bilinear form.
This paper is organized as follows: 
In section \ref{sec:BBS}, we give a brief introduction of the BBS,
  which is compared with the ffBBS we propose in section \ref{sec:ffBBS}.
In section \ref{sec:p3}, the ffBBS with $p=3$ is shown together
  with numerical simulations that exhibit stable propagations of solitons over finite field $\F_3$.
Then in section \ref{sec:proof}, the exact one-soliton solutions, 
  which have a nested structure similar to fractals,  are proved.
The last section \ref{sec:remarks} is devoted to the conclusion and summary obtained in this paper. 

\section{Box-Ball System (BBS)}
\label{sec:BBS}

In this paper we propose the system that is an analogue of the BBS and investigate its properties.
For the purpose of comparison, we summarize several matters relating to the BBS below
\cite{TS90, TTMS, TTM00}.

\begin{figure}[bhtp]
  \begin{center}
    \includegraphics[width=100mm]{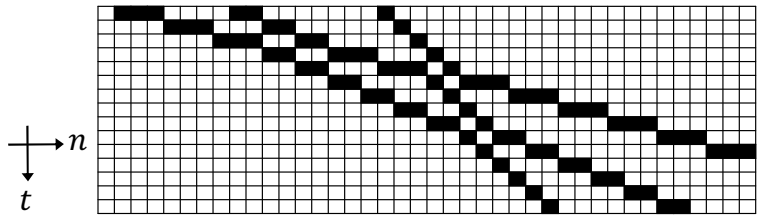} 
    \caption{A example of time evolution of BBS by (\ref{eq:BBS}) with $L=1$.}
    \label{fig:BBSsoliton1}
  \end{center}
\end{figure}

The BBS in the case that the capacity of each box is constant $L$ is shown as follows:
\begin{equation}
  u^{t+1}_{n} := \min
  \left\{
    L-u^{t}_{n}, \sum^{n-1}_{i=-\infty} \left( u^{t}_{i} - u^{t+1}_{i} \right)
  \right\} + \max (0, -L), 
\label{eq:BBS}
\end{equation}
where $u^{t}_{n}$ is the dependent variable that is interpreted as
  the number of balls in the box at time $t$ and position $n$.
By the transformation between the dependent variables
\begin{equation}
  u^{t}_{n} = s^{t}_{n} - s^{t}_{n-1}, \qquad
  s^{t}_{n} = g^{t-1}_{n} - g^{t}_{n}, 
  \label{eq:usg}
\end{equation}
we obtain the ultradiscrete {\it bilinear} form under appropriate boundary conditions as follows:
\begin{eqnarray}
  s^{t+1}_{n+1} - s^{t}_{n} - \max (0, -L) & = & \min \left( 0, L - s^{t}_{n+1}+s^{t+1}_{n} \right),
  \label{eq:udS} \\
  g^{t+1}_{n+1} + g^{t-1}_{n} + \max(0, -L) & = & \max \left( g^{t}_{n+1}+g^{t}_{n}, g^{t-1}_{n+1}+g^{t+1}_{n} - L \right).
  \label{eq:udBilinear}
\end{eqnarray}
Because the parameter $L$ is usually interpreted as the maximum number of balls
  that are stuffed into each box, $L$ is assumed to be positive ($L>0$) for obtaining meaningful equations
  as the dynamics of ball moving. 
Ordinarily, the term $\max(0, -L)$ is eliminated as a matter of course. 
However, the term $\max(0, -L)$ is left in (\ref{eq:BBS})--(\ref{eq:udBilinear})
  as it is for the next section on purpose.
Furthermore by the transformation of variables $c^{t}_{n} := s^{t}_{n-1} - s^{t+1}_{n-1}$, 
  we obtain
\begin{eqnarray}
  u^{t+1}_{n} 
   & = & \min \left( L-u^{t}_{n}, c^{t}_{n} \right) + \max (0, -L) \nonumber \\
   & = & c^{t}_{n} - \max \left( u^{t}_{n}+c^{t}_{n}-L, 0 \right) + \max (0, -L), 
\label{eq:BBSCInfty1} \\
   c^{t}_{n+1}
   & = & u^{t}_{n} + c^{t}_{n} - u^{t+1}_{n} \nonumber \\
   & = & u^{t}_{n} + \max \left( u^{t}_{n}+c^{t}_{n}-L, 0 \right) - \max (0, -L)
\label{eq:BBSCInfty2}
\end{eqnarray}
from (\ref{eq:udS}), which stands for the combinatorial $R$\cite{Hikami99}. 
This dependent variable $\left\{ c^{t}_{n}\right\}$ is called a carrier\cite{TM97}.
The BBS we have shown here is the case of the carrier without limit
  (the capacity of the carrier is $\infty$).
The time evolution of the BBS from $t$ to $t+1$ with $L=1$ is interpreted as the 
interaction between this carrier and the balls:
\begin{enumerate}
\item The carrier repeats (ii) and (iii) through all boxes from left to right.
\item If a ball is in the box, the carrier picks up the ball.
\item If the box is empty and the carrier contains at least one ball, the carrier drops one ball into the box.
\end{enumerate}
Figure \ref{fig:BBSsoliton1} shows the examples of time evolution of the BBS.

\section{Soliton equation over finite fields (ffBBS)}
\label{sec:ffBBS}

The BBS we have observed in the previous section is the well-studied system, and its important
  properties such as solutions, conserved quantities etc.~are well-known.
It is genuine integrable soliton systems
  with the $\left\{0, 1, \ldots, L \right\}$-valued dependent variables $\left\{ u^{t}_{n} \right\}$. 
However, even in the case $L=1$, 
  (\ref{eq:BBS})--(\ref{eq:udBilinear}) are equations over not binary but integer (or real number), 
  because the state of the carrier $c^{t}_{n}$, i.e. the number of balls contained in the carrier,
  is specified by an integer (or real number). 
One of our motivation stated in the introduction is to obtain
  an integrable system which does not have relation to ultradiscretization.
In the following section, we define a system of which all dependent variables are the elements in finite set,
  which also have good algebraic structures, namely the finite fields.

\subsection{A system similar to BBS}
In this paper, we consider an analogue of the BBS over finite fields, based on 
  the expressions (\ref{eq:BBS})--(\ref{eq:udBilinear}).
However, for example, trials for regarding (\ref{eq:BBS}) as an equation over finite fields 
  do not apparently go well.
Because the finite fields are not totally ordered sets and there is no definitions of 
  magnitude relationship, we may not apply the minimum function $\min(\cdot, \cdot)$ in (\ref{eq:BBS}) as it is.
To that end, we observe the algebraic operation needed for the transformations among equations
  (\ref{eq:BBS})--(\ref{eq:udBilinear}).
Then we find that, in addition to the four arithmetic operations of finite fields, the distributive law for
  the maximum function as binary operation, that is, 
\begin{equation}
\max (x, y) + z = \max (x+z, y+z)
\label{eq:distributiveLaw}
\end{equation}
 are used ($x, y, z \in \R $).
Note that the addition is `$\max$' and the multiplication `$+$' in max-plus algebra, 
  and that $\min(\cdot, \cdot)$ is defined by $\max(\cdot, \cdot)$ as $\min (x, y) := -\max (-x, -y)$.
In other words, the operation needed for $\max$ in the transformations 
  (\ref{eq:BBS})--(\ref{eq:udBilinear}) is only the distributive law 
  for $\max$ and $+$, and such a property that $\max$ is the function that returns the maximum value 
  is not required.

Accordingly, we make a trial of replacing the binary operation $\max : \R \times \R \to \R$ by a function  
$M : \F_q \times \F_q \to \F_q $ with the following properties (\ref{eq:P0}), (\ref{eq:P1}), and (\ref{eq:P2}); 
\begin{eqnarray}
M(0, 0) = 0, & & \label{eq:P0} \\
M(a, b) = M(b, a)  & : & \mbox{commutative law} \label{eq:P1} , \\
M(a, b) + c = M(a+c, b+c) \quad & : & \mbox{distributive law} \label{eq:P2}, 
\end{eqnarray}
where $a, b, c \in \F_q$ and $q := p^m$. 
These properties yield 
\[
  M(a, b) - M(-a, -b) =  a + b, 
\]
  which may be regarded as an analogue of the formula of $\max$
\[
  \max (x, y) - \max(-x, -y) = \max(x, y)+\min(x, y) = x+y.
\]
Namely, $\min(\cdot, \cdot)$ in (\ref{eq:BBS}) is also replaced by means of this function $M$ as above.
\begin{Proposition}
In the case $p=2$, there does not exist a function $M$ that satisfies (\ref{eq:P0}), (\ref{eq:P1}), and (\ref{eq:P2}).
\label{prop:odd}
\end{Proposition}
\begin{Proof}
$ M(0, 1) = M(-1, 0) + 1 = M(1, 0) + 1 = M(0,1) + 1 $.
\qed
\end{Proof}
\begin{Proposition}
If the associative law
\begin{eqnarray}
  M(a, M(b, c)) = M(M(a, b), c) \label{eq:P3} 
\end{eqnarray}
  is also required in addition with (\ref{eq:P0}), (\ref{eq:P1}), and (\ref{eq:P2}), 
  there does not exist a binary operation $M$ over ${\F_{q}}^{2}$
  that satisfies these four properties.
\footnote{
The associative law is needed for the $N$-soliton solutions of the original BBS to satisfy (\ref{eq:BBS}).
Therefore the known solutions of BBS cannot be applied to construct solutions of (\ref{eq:ffBilinear}).
}
\end{Proposition}
\begin{Proof}
We will prove by contradiction 
  with the assumption that there exists $M$ that satisfies (\ref{eq:P0}), (\ref{eq:P1}), (\ref{eq:P2}), and (\ref{eq:P3}). 
In this proof, define $f(x) := M(0,x)$ and ${\cal I} := \left\{ f(x) \left| x \in \F_q \right. \right\} \subseteq \F_q$, 
  which is the image of the map $f$. 
Then, $f(0)=0$ immediately follows from (\ref{eq:P0}), and $\forall x \in \F_q, f(x)=f(-x)+x$ from $M(x, 0) = M(0, -x) + x$, 
  and $\forall x \in \F_q, f(f(x)) = f(x)$ from $M(M(x, 0), 0) = M(x, M(0, 0)) = M(x, 0)$.
That is, for all $y \in {\cal I}$, $f(y)=y$ and therefore $f(-y) = f(y)-y = 0$.
Note that $0 \in {\cal I}$ from $f(0)=0$.
For all $y, z \in {\cal I}$, we also obtain the following: 
\begin{eqnarray*}
  f(y+z)
  & = & f(f(y)+z) \\
  & = & M(M(y, 0)+z, 0) \\
  & = & M(M(y, 0), -z) + z \\
  & = & M(y, M(0, -z)) + z \\
  & = & M(0, M(0, -z)-y) + y + z \\
  & = & f(f(-z)-y) + y + z \\
  & = & f(-y) + y + z \\
  & = & y + z. 
\end{eqnarray*}
If ${\cal I} = \left\{ 0 \right\}$, then since ${\cal I}$ is the image of the map,
  the function values are always zero $f(x) = 0$ and $f(-x) = 0$
  even for non-zero $x \in \F_q^{\times}$. This contradicts as $x = f(x)-f(-x) = 0$. 
Thus, the image ${\cal I}$ contains at least one non-zero element $\exists x_{*} (\neq 0) \in {\cal I}$ 
  and is at least one or higher-dimensional vector space with the non-zero element $x_{*}$.

For this element $x_{*} \in {\cal I}$, we have $f(-x_{*})=0$ whereas 
  $f(-x_{*}) = f((p-1)x_{*}) = (p-1)f(x_{*}) = -f(x_{*}) = -x_{*}$, where $p$ is the characteristic of $\F_q$ ($q=p^m$).
This contradicts $x_{*} \neq 0$, so the proof is complete.
\qed
\end{Proof}

Let $\nu$ be $(p-1)/2 \in \Z$ because $p$ is an odd prime from proposition \ref{prop:odd}. 
For simplicity, we restrict ourselves to the prime field case ($m=1$) hereafter. 
The map $M$ that satisfies (\ref{eq:P0}), (\ref{eq:P1}), and (\ref{eq:P2}) is uniquely parametrized
by only the values of $M(0, j) \ (j=1, 2, \ldots, \nu)$, 
because $M(0, -j)$ is obtained from the relation
$M(0, -j)= M(0, j)-j$ by noticing $-j=\nu+1, \nu+2, \ldots, 2\nu=p-1$.
Hence, the $M$ may be expressed by $\nu$-tuple of $e_j \left(\equiv M(0, j) \right)$ as
\[
M(0, a) \equiv M^{(e_1, e_2, \ldots, e_\nu)}(0, a) := \sum^{\nu}_{j=1} (\delta_{a, j} e_j + \delta_{a, -j} (e_j -j)), 
\]
where $\delta$ is the Kronecker delta, 
\begin{eqnarray*}
  \delta_{b+c, c} & := & -(b+1)(b+2)\cdots (b+p-1) \\
  & = & 
  \left\{
  \begin{array}{ll}
    1 & (b=0) \\
    0 & \mbox{otherwise}
  \end{array}
  \right. , \ 
  \forall b, c \in \F_p, 
\end{eqnarray*}
which is equivalent to Wilson's Theorem; $(p-1)!+1 \equiv 0 \  (\mbox{mod} \ p )$.
Note that $M(0, a)$ is the polynomial, which is parametrized by $(e_1, e_2, \ldots, e_\nu) \in {\F_p}^{\nu}$.
We will abbreviate $M^{(e_1, e_2, \ldots, e_\nu)}$ to $M$ hereafter. 

By means of this polynomial function $M$, we propose the following {\it bilinear} equation
  as the starting point of our system over finite fields (cf.~(\ref{eq:udBilinear})). 
%\footnote{(\ref{eq:ffBilinear}) is called {\it bilinear} by imitating the (ultra-)discrete systems, though it is already not the case.}
\begin{equation}
  G^{t+1}_{n+1} + G^{t-1}_{n} + M(0, -L) =
  M \left( G^{t}_{n+1}+G^{t}_{n}, G^{t-1}_{n+1}+G^{t+1}_{n} - L \right).
\label{eq:ffBilinear}
\end{equation}
We will refer to (\ref{eq:ffBilinear}) as the box-ball system over finite fields (ffBBS). 
Introducing the dependent variables similar to (\ref{eq:usg}) as 
\begin{equation}
  U^{t}_{n} = S^{t}_{n} - S^{t}_{n-1}, \qquad
  S^{t}_{n} = G^{t-1}_{n} - G^{t}_{n}, 
  \label{eq:USG}
\end{equation}
we obtain
\begin{equation}
  U^{t+1}_{n} = -M
 \left(
   -L+U^{t}_{n}, -\sum^{n-1}_{i=-\infty} \left( U^{t}_{i} - U^{t+1}_{i} \right)
\right)  + M(0, -L), 
\label{eq:ffBBS}
\end{equation}
\begin{equation}
  S^{t+1}_{n+1} - S^{t}_{n} -M \left( 0, -L \right) =  -M \left( 0, -L + S^{t}_{n+1}-S^{t+1}_{n} \right), 
\label{eq:ffS}
\end{equation}
which are the analogues of (\ref{eq:BBS})--(\ref{eq:udBilinear}).
The dependent variables $G^{t}_{n}$, $S^{t}_{n}$, $U^{t}_{n}$, and the parameter $L$ are elements in $\F_p$,
  and the independent variables $n$ and $t$ are those in $\Z$.
The carrier interpretation (\ref{eq:BBSCInfty1}) and (\ref{eq:BBSCInfty2}) also holds by replacing $\max$ by $M$.
This ffBBS is apparently reversible with respect to $t \to -t$ and $n \to -n$, 
  which follows from the so-called unitarity condition of $R$ matrix. 
Note that the indeterminate such as division by zero does not appear for all initial values
  because the evolution in this system is described by polynomials.
This is a remarkable feature of (\ref{eq:ffBilinear})-(\ref{eq:ffS}) compared with
  the preceding studies of dynamical systems over finite fields stated in the introduction.

\subsection{zero-soliton solution}
In BBS (\ref{eq:udBilinear}), it is well known that $g^{t}_{n}=\mbox{const.}$ gives the zero-soliton solution
  and yields $u^{t}_{n} = 0$, i.e., the state without balls. 
From the above correspondence, it is clear that $G^{t}_{n}=\mbox{const.}$ gives the zero-soliton solution
  and yields $U^{t}_{n} = 0$ for ffBBS (\ref{eq:ffBBS}). 
Thus, we may easily obtain the zero-soliton solution due to the formal analogue. 
In the next section, we will observe the typical solutions of the ffBBS. 
\section{Simplest case: $p=3$}
\label{sec:p3}

Hereafter we limit ourselves to the prime field $\F_3$, which is the simplest case 
  because $p$ should be odd from proposition \ref{prop:odd}. 
In this case $p=3$,  the function $M$ is determined by only one parameter $e_1 \in \F_3$ due to $\nu = 1$ as
\begin{equation}
  M^{(e_1)}(a, b) := (e_1+1) (a^2+a b+b^2) + 2 (a+b).
  \label{eq:Ma}
\end{equation}

For $e_1 = 2$, the function $M^{(2)}$ become first-degree polynomial and determine only trivial systems. 
The relation $M^{(1)}(a, b)  = 2^{-1} M^{(0)}( 2 a, 2 b)$ implies that
  $M^{(1)}$ and $M^{(0)}$ are isomorphic ($2^{-1} \equiv 2 \bmod 3 $). 
For this reason, we hereafter consider only $M^{(1)}(a, b) = 2(a^2+a b+b^2 + a+b)$ as $M$.
The following is then satisfied. 
\begin{Fact}
  $M^{(1)}(0, 0)=0, M^{(1)}(0, 1)=1, M^{(1)}(0, 2)=0$.
  \label{fact:M}
\end{Fact}

Let $L$ be fixed to one so that $M(0, -L) = 0$ for simplicity. 
In this paper we will obtain the one-soliton solutions only for this particular $\F_3$ with $e_1=1$ and $L=1$.
The general $\F_q$ cases are subjects for future analysis, though our numerical experiments
  show abundant solitonic patterns. 

\subsection{One-soliton solutions with integer velocities}
\begin{figure}[bthp]
  \vspace*{5mm}
  \begin{center}
    \includegraphics[width=120mm]{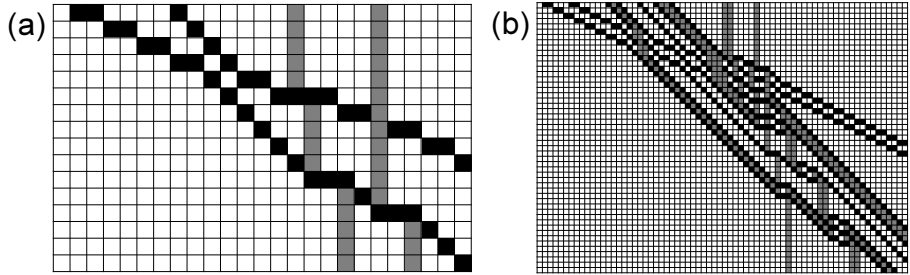}
    \caption{
      Examples of time evolution of ffBBS by (\ref{eq:ffBBS}) with $p=3, e_1 = 1, L=1$ \ (White: 0, Black: 1, Gray: 2).
      As initial values, solitons which have velocities (a) 2, 1, 0, 0 (b) 2, 2, 1, 1, 1, 0, 0 are given at uppermost row.
    }
    \label{fig:ffBBS0102}
  \end{center}
\end{figure}

The stable solitary wave that is separated by sufficient `0's is called {\it soliton} in this paper. 
An example of three kinds of solitons is shown in figure \ref{fig:ffBBS0102}(a) ($p=3, e_1 = 1, L=1$).
First, the pattern with `0's (e.g.~white part in lower left area) is zero-soliton solutions. 
On this background, the solitons `11', `1', `2', `2' are moving to the right with 
  the velocity 2, 1, 0, 0, respectively.
Stable propagations are observed before and after collisions
  with some phase shifts, which are similar to the BBS solutions in figure \ref{fig:BBSsoliton1}. 

We elucidate the travelling waves by assuming $G(\xi) \equiv G^{t}_{n}$, $\xi := K n - \Om t$.
Substituting these into (\ref{eq:ffBilinear}) and (\ref{eq:ffS}), we obtain
\begin{eqnarray}
  G(\xi+K-\Om)+G(\xi+\Om)  & = & \nonumber \\
    & & \hspace*{-35mm} M \left( G(\xi+K)+G(\xi), G(\xi+K+\Om)+G(\xi-\Om)-1 \right),
    \label{eq:travelG}
\end{eqnarray}
\begin{equation}
  S(\xi+K-\Om)-S(\xi) = -M \left( 0, S(\xi+K)-S(\xi-\Om)-1 \right).
  \label{eq:travelS}
\end{equation}
Choosing $\Om = 0$ yields the solution with velocity zero $S(\xi+K)-S(\xi)=0 \mbox{ or } 2$
  from (\ref{eq:travelS}); that is, $U^{t}_{n+1}=0 \mbox{ or } 2$,
  since $U^{t}_{n+1}=S(\xi+K)-S(\xi)$.
Thus initial states that consist of only `0's and `2's give the patterns that do not move. 

Next, we consider travelling waves with velocity one.
Applying $\Om = K$ to (\ref{eq:travelG}) with the distributive law (\ref{eq:P2}) yields
%\begin{equation*}
%  G(\xi)+G(\xi+K) = M \left( G(\xi+K)+G(\xi), G(\xi+2K)+G(\xi-K)-1 \right), 
%\end{equation*}
\begin{equation}
   M \left( 0, G(\xi+2K)-G(\xi+K)-G(\xi)+G(\xi-K)-1 \right) = 0.
  \label{eq:MG4}
\end{equation}
Substituting  
\begin{eqnarray*}
U^{t}_{n} & = & G^{t-1}_{n}-G^{t}_{n}-G^{t-1}_{n-1}+G^{t}_{n-1} \\
 & = & G(\xi+\Om)-G(\xi)-G(\xi-K+\Om)+G(\xi-K) \\
 & = & G(\xi+K)-2 G(\xi)+G(\xi-K)
\end{eqnarray*}
into (\ref{eq:MG4}), we obtain the implicit equation
$
   M \left( 0, U^{t}_{n+1} + U^{t}_{n} -1  \right) = 0.
$
The solution is $U^{t}_{n+1} + U^{t}_{n} = 0 \mbox{ or } 1$; that is, 
\begin{equation}
  \ds (U^{t}_{n}, U^{t}_{n+1}) \in \left\{ (0, 0), (0, 1), (1, 0), (1, 2), (2, 1), (2, 2) \right\}
  \label{eq:velocity1}
\end{equation}
for all $t$ and $n \in \Z$.
Since the zero-soliton solution consists of only `0's and the combination $(0, 2)$ is not contained in the above set,
  we may find that the left end of the pattern of the travelling wave with velocity one starts from `1'.
Likewise, we may find that the right end of the pattern ends with `1'
  since $(2, 0)$ is not allowed. 
Shifting $n$ and combining  the adjacent $U^{t}_{n}$ and $U^{t}_{n+1}$ in (\ref{eq:velocity1}), 
  we obtain the one-soliton patterns 
  `$\cdots 010 \cdots$' in the case without `$2$', 
  and `$\cdots 012(12+2)^{*}10 \cdots $' (regular expression) otherwise.
(e.g.~`010', `01210', `012210', `0121210', `0122210', $\dots$ ).

Finally, we consider travelling waves with velocity two.
Applying $\Om = 2 K$ to (\ref{eq:travelG}) yields
\begin{eqnarray*}
  G(\xi-K)+G(\xi+2K) & = & \\
  & & \hspace*{-35mm} M \left( G(\xi+K)+G(\xi), G(\xi+3K)+G(\xi-2K)-1 \right).
\end{eqnarray*}
Substituting $U^{t}_{n} = G(\xi+2K)-G(\xi+K)-G(\xi)+G(\xi-K) $ into the above equation, 
  we obtain $U^{t}_{n} = M \left( 0, U^{t}_{n+1} + U^{t}_{n}+ U^{t}_{n-1} -1  \right)$, 
  which have the solutions
\begin{equation*}
  \ds (U^{t}_{n-1}, U^{t}_{n}, U^{t}_{n+1}) \in \left\{ (0, 0, 0), (0, 0, 1), (0, 1, 1), (1, 0, 0), (1, 1, 0) \right\}
\end{equation*}
for all $t$ and $n \in \Z$. 
Combining these adjacent relations, we obtain only `$\cdots$001100$\cdots$'. 

Figure \ref{fig:ffBBS0102}(b) shows the examples of solitons that we have proved here.
Note that we regard the pattern with only 0's (white cells) as the zero-soliton solution rather than the one with velocity zero.
Each shape of the solitons is preserved as well in spite of collisions. 
We summarize the above solutions as follows:
\begin{Theorem}[one-soliton solutions with integer velocities]
\begin{itemize}
\item The patterns with velocity zero consist of only `$2$'.
\item The patterns with velocity one are, `$010$' and `$012(12+2)^{*}10$' (regular expression). 
Namely, there exist infinite kinds of patterns with velocity one that does not contain `$0$' inside them
  (`$010$', `$01210$', `$012210$', `$0121210$', `$0122210$', $\dots$ ).
\item The pattern with velocity two is only `$11$'.
\end{itemize}
\label{thm:easy}
\end{Theorem}

Besides the soliton solutions with integer speed 0, 1, and 2 shown in theorem \ref{thm:easy},
there exist travelling waves with $\Om = K+1 =2^{h+1}$ for all $h \in \Z_{>0}$.
These solutions are shown in theorem \ref{thm:concrete}, 
  which contain the pattern `$12^h0^h1$' at a certain time.
In the case $h=0$, this pattern coincides with the soliton `11' in theorem \ref{thm:easy}.

\begin{figure}[bt]
  \begin{center}
    \includegraphics[width=10cm]{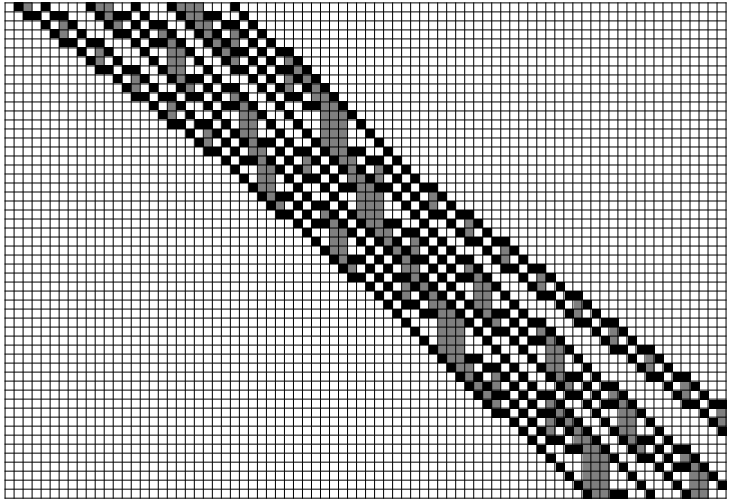} 
    \caption{
      An example of time evolution by (\ref{eq:ffBBS}) with $p=3$, $e_1 = 1$, and $L=1$ \ (White: 0, Black: 1, Gray: 2).
      As initial values, three solitons that have velocity 4/3, 8/7, 16/15 respectively 
      described in theorem \ref{thm:concrete} are given at uppermost row. 
      Despite complex collisions, three solitons conserve their patterns. 
    }
    \label{fig:ffBBS03}
  \end{center}
\end{figure}

Figure \ref{fig:ffBBS03} shows an example of solitonic interactions between these travelling waves, 
  in which the patterns `1201', `122001', `12220001' are given respectively as initial values. 
The dispersion relation $\Om = 2^{h+1}$, $K =2^{h+1}-1$ leads to the fractional velocities
$2^{h+1}/(2^{h+1}-1)$; that is, the same pattern appears again at $2^{h+1}-1$ time later
at the position shifted $2^{h+1}$.
Note that (\ref{eq:ffBBS}) is reversible.

It is not proved whether there exist one-soliton solutions
  besides those shown in this section.
However, our numerical experiments show that the above patterns are all.

\section{One-soliton solutions with fractional velocities}
\label{sec:proof}

In this section, we prove that there exist the travelling waves with the fractional 
velocities $2^{h+1} / (2^{h+1}-1)$ for all $h \in \Z_{>0}$.

\subsection{Definitions}

Let $h$ be a positive integer and  $\Om = K+1 =2^{h+1}$ through this section. 
Suppose the integer sequences that are recursively defined as follows:
\begin{itemize}
\item $I^{(h)}_{h} := ( 2^{h+1}-1 )$.
\item For \mbox{$1 \leq i \leq h$}, the sequence \mbox{$I^{(h)}_{i-1}$} is defined by the following two steps: 
  \begin{enumerate}
    \item copy $I^{(h)}_{i}$.
    \item simultaneously replace all $\underline{2^{i+1}-1}$ with $\underline{( 1, 2^{i}-1, 2^{i}-1)}$.
  \end{enumerate}
\end{itemize}
Note that $I^{(h)}_{i}$ ($1 \leq i \leq h$) consists of $1$'s and $(2^{i+1}-1)$'s, and $1$'s are never replaced in the step (ii).
Though we represent the sequences with parentheses to express the tree structure 
  accompanied by the replacements of integers,
  the parentheses are ignored as for the integer sequences (cf.~figure \ref{fig:tree}).
For specification of each element in $I^{(h)}_{i}$, we define the numbering 
\[
I^{(h)}_{i} =: ( I^{(h)}_{i}(0), I^{(h)}_{i}(1), \ldots, I^{(h)}_{i}(2^{h-i+1}-2) ),
\]
where the length of the integer sequence $I^{(h)}_{i}$ is $2^{h-i+1}-1$.
The summation of the elements in this sequence $I^{(h)}_{i}$ is
\begin{equation}
  \sum_{k=0}^{2^{h-i+1}-2} I^{(h)}_{i} (k) = 2^{h+1}-1 = K \quad (0 \leq i \leq h),
  \label{eq:sum_ini} 
\end{equation}
  and does not depend on $i$
  because the two underlined parts above have the same summation $2^{i+1}-1$.

We define the palindromic sequence $I^{(h)}$ by aligning $I^{(h)}_{i}$ as follows:
\begin{eqnarray}
  I^{(h)} & := &
   ( I^{(h)}_{h}, I^{(h)}_{h-1}, \ldots, I^{(h)}_{1}, \overbrace{1, 1, \ldots, 1}^{2^{h+1}+h-1}, I^{(h)R}_{1}, I^{(h)R}_{2}, \ldots, I^{(h)R}_{h} )
  \label{eq:Iseq} \\
  & = & 
   ( I^{(h)}_{h}, I^{(h)}_{h-1}, \ldots, I^{(h)}_{1}, I^{(h)}_{0}, \overbrace{1, 1, \ldots, 1}^{h}, I^{(h)R}_{1}, I^{(h)R}_{2}, \ldots, I^{(h)R}_{h} ),
  \label{eq:IseqwithI0} 
\end{eqnarray}
where $X^R$ is the reverse string of $X$ (each of symbol here is integer). Simple calculations show that
  the length of the sequence $I^{(h)}$ is $r:=3 \cdot 2^{h+1}-(h+5)$, and the summation is $(2h+1)2^{h+1}-(h+1)$.
Note that (\ref{eq:IseqwithI0}) follows from the fact that $I^{(h)}_{0}$ consists of $(2^{h+1}-1)$ 1's.

Furthermore let us define the sequential numbering of the elements in $I^{(h)}$ with the length $r$ as 
$(b_0, b_1, \ldots, b_{r-1} ) := I^{(h)}$. In other words, for $0 \leq \forall l \leq 2^{h+2}-(h+4)$,
there exists a unique $i$ such that $2^{h-i+1}-(h-i+2) \leq l < 2^{h-i+2}-(h-i+3)$, and we have
\begin{equation}
  0 \leq i \leq h, \ 
  b_{l} = I^{(h)}_i (l-2^{h-i+1}+(h-i+2)).
  \label{eq:defb}
\end{equation}
For $2^{h+2}-(h+3) \leq l \leq r-1 $, $b_{l}$ may be palindromically defined by $b_{r-l-1}$.
We also define the set $J := \left\{ a_0, a_1, \ldots, a_r \right\}$ by
  the recurrence equation $a_{l+1} = a_l + b_l, a_0=0$. 
Namely, $a_l$ is the summation of $b_l$, and $a_{r} = (2h+1)2^{h+1}-(h+1)$ is the maximal element in $J$. 
Suppose the function $S(\xi) := \ds \sum_{\mu \in J} \theta(\xi - \mu) \bmod 3 \in \F_3 $,
  where $\theta(\xi)$ be the unit step function that is defined over integer $\xi$ as
$ \ds
  \theta ( \xi ) =   \left\{
  \begin{array}{ll}
    1 & ( 0 \leq \xi ) \\
    0 & \mbox{otherwise}
  \end{array} 
  \right.
$.
We then obtain the following theorem.
\begin{Theorem}
Let $h$ be an positive integer, and $\Om = K+1 = 2^{h+1}$.
Then, $S(\xi)$ defined above is the solution of (\ref{eq:travelS}).
The corresponding travelling wave solution $\left\{ U^{t}_{n} \right\}$ of (\ref{eq:ffBBS}) contains the pattern `$12^h0^h1$' 
  at a certain time, and have fractional velocity
%  $\ds \frac{\Om}{K} = \frac{2^{h+1}}{2^{h+1}-1}$.
  $\Om / K = 2^{h+1}/(2^{h+1}-1)$.
\label{thm:concrete}
\end{Theorem}

\begin{Example}[The case $h=3$]
  \normalfont
Suppose $h=3$ as an example, then we have $K=15, \Om=16, r=40$.
The initial sequence is defined as $I^{(3)}_{3} = ( 15 )$ with the length one. 
$I^{(3)}_{2}$ is obtained by replacing the value 15 in $I^{(3)}_{3}$ with $(1, 7, 7 )$,
  i.e., $I^{(3)}_{2} = ( (1, 7, 7 ) )$.
Furthermore, replacing each 7's in $I^{(3)}_{2}$ with $(1, 3, 3 )$ respectively yields 
  $ I^{(3)}_{1} = ( ( 1, (1, 3, 3), (1, 3, 3) ) )$.
Combining these sequences, we obtain
\begin{eqnarray*}
  I^{(3)} & = & ( ( 15 ), ( (1, 7, 7) ), (  ( 1, (1, 3, 3), (1, 3, 3) ) ), \\
  & & \qquad \overbrace{ 1, 1, 1, 1, 1, 1, 1, 1, 1, 1, 1, 1, 1, 1, 1, 1, 1, 1}^{18}, \\
  & & \qquad ( ( ( 3, 3, 1), (3, 3, 1), 1 ) ), ( (7, 7, 1 ) ), ( 15 ) ), 
\end{eqnarray*}
and 
$b_0 = 15, b_1 = 1, b_2 = 7, b_3 = 7, b_4 = 1, \ldots, b_{38} = 1, b_{39} = 15 $, and
$a_0 = 0, a_1 = 15, a_2 = 16, a_3 = 23, a_4 = 30, a_5 = 31, \ldots, a_{39} = 93, a_{40} = 108 $ sequentially from the left. 
Substituting $K$ and $\Om$ into (\ref{eq:travelS}) yields 
\[
  S(\xi)-S(\xi-1) = M \left( 0, S(\xi+15)-S(\xi-16)-1 \right).
\]
Let us suppose $\xi=15$ for example, then this equation holds as follows.
The left-hand side of the above equation equals one due to $a_1=15$, 
  while the right-hand side results in $M(0, 5-1) = M(0, 1) = 1$
  because the elements in $J$ that satisfy $0\leq a_i \leq 30$ are the ones with $0 \leq i \leq 4$ and accordingly
  the number of them is five. 
\label{ex:n3}
\end{Example}

The rewriting rule in the above definition of $I^{(h)}_{i-1}$ is simultaneously applied. 
Therefore this grammar is a kind of L-system (Lindenmayer system)\cite{LSYSTEM}. 
Consider the grammar $G' := \langle \Sigma', \om, P \rangle $, where 
  the alphabet $\Sigma' = \left\{ {\tt A}, {\tt B} \right\}$,
  the start symbol $\om = {\tt B} \in \Sigma'$,
  the production rule $P=\left\{ {\tt A} \to {\tt A}, {\tt B} \to {\tt A}{\tt B}{\tt B} \right\} $.
Replacing the derived strings 
  $\left\{ {\tt B}, {\tt A}{\tt B}{\tt B}, {\tt A}{\tt A}{\tt B}{\tt B}{\tt A}{\tt B}{\tt B}, \ldots \right\}$
  with ${\tt A} \mapsto 1, {\tt B} \mapsto 2^{i}-1$ then yields $I^{(\cdot)}_{i-1}$. 
Thus, the nested structure in our solutions is obvious from this grammar.

In the next subsection, the sequence $I^{(h)}_{i}$ are associated with a positional numerical system
  so that we may obtain the value $b_l$ from the index $l$.

\subsection{Positional numerical system with mixed radix}

In this section, we prepare a positional numerical system with mixed radix. 
Suppose that an integer $i$ such that $0 \leq i \leq 2^{m+1}-2$ for given $m \in \Z_{\geq 1}$.
Let the number $i$ be represented by
  $i = \ds \sum_{k=1}^{m} d_k (2^k -1)$, $d_k \in \left\{ 0, 1, 2 \right\} \subset \Z $,
  provided that if there exists $k'$ such that $d_{k'}=2$, then $d_{k'-1} = d_{k'-2}= \cdots = d_{1} = 0$.
Namely, $d_k$ is the numerals in $\left\{ 0, 1, 2 \right\}$ and the corresponding radix for $d_k$ is
$2^k-1$ depending on $k$.
As examples, 
\begin{itemize}
\item In the case $m=1$, $d_1=0, 1, 2$ for $0\leq i \leq 2 $ respectively,
\item In the case $m=2$,  $(d_2, d_1)=(0, 0), (0, 1), (0, 2), (1, 0), (1, 1), (1, 2), (2, 0) $ 
  for $0\leq i \leq 6$ respectively. Note that $(2, 1)$ and $(2, 2)$ are forbidden by the above restriction.
\end{itemize}
For any given $m$, we show that this positional notation is not redundant in proposition \ref{prop:bijection}. 
In this paper, we do not omit leading zeros unlike usual positional notations.

\begin{Definition}
Let $m$ be a positive integer and define the following formal language $\calL_m$
over the alphabet $\Sigma = \left\{ \tt0, \tt1, \tt2 \right\} $:
\begin{eqnarray*}
\ds \calL_m & := & \left\{ d_m d_{m-1} \cdots d_2 d_1 \in \Sigma^m \left. \right|
  \ 1 \leq \forall i \leq m, d_i \in \Sigma, \right. \mbox{if there exists $k'$ }
\\
  & & \hspace*{2mm} \left.
  \mbox{such that $d_{k'}={\tt 2}$, then $d_{k'-1} = d_{k'-2}= \cdots = d_{1} = {\tt 0}$}
  \right\}, 
\end{eqnarray*}
where $\Sigma^m := \Sigma^{m-1} \times \Sigma$ with multiplication as 
  the concatenation of strings, and 
$\Sigma^{0} := \left\{ \varepsilon \right\} $,
$\varepsilon$ the empty string.
We also define $\calL_0 := \Sigma^{0}$ and $\calL := \bigcup_{m=0}^{\infty} \calL_{m}$.

For $m \in \Z_{\geq 0}$, we denote the concatenation of the strings $\beta \in \Sigma^m$
  and the alphabet $a \in \Sigma$ by $\beta a \in \Sigma^{m+1}$, 
  and conversely the operation that eliminates one character from the tail of
  $\beta a \in \Sigma^{m+1}$ as
  $\mbox{del} : \Sigma^{m+1} \to \Sigma^m, \beta a \mapsto \beta$.
\end{Definition}

Hereafter we identify the element in $\Sigma=\left\{ {\tt 0}, {\tt 1}, {\tt 2} \right\}$
  with that in the set $\left\{ 0, 1, 2 \right\} \subset \Z$ respectively.

\begin{Proposition}
Let $m$ be a positive integer. 
Then the map 
$ 
  C_{m} : \calL_m \to \left\{ 0, 1, \ldots, 2^{m+1}-2 \right\} \subset \Z, 
  d_m d_{m-1} \cdots d_1 \mapsto \sum_{k=1}^{m} d_k (2^k -1)
$
%\begin{eqnarray*}
%  C_{m} & : & \calL_m \to \left\{ 0, 1, \ldots, 2^{m+1}-2 \right\} \subset \Z, \\
%  & & d_m d_{m-1} \cdots d_1 \mapsto \sum_{k=1}^{m} d_k (2^k -1)
%\end{eqnarray*}
is bijective.
The map $C_0$ is defined as $C_0 : \varepsilon \mapsto 0 $, which is also bijective.
\label{prop:bijection}
\end{Proposition}

\begin{figure}[btp]
  \begin{center}
    \includegraphics[width=40mm]{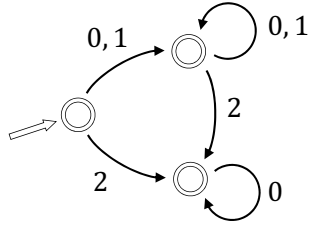} 
    \caption{Finite automaton that accepts the language $\calL$.}
    \label{fig:FA}
  \end{center}
\end{figure}

\begin{Proof}
  \normalfont 
If $m=1$, the identification ${\tt 0} \equiv 0, {\tt 1} \equiv 1, {\tt 2} \equiv 2$ is one-to-one correspondence.
Now assume the map $C_m$ is bijective for induction.
For $0 \leq i \leq 2^{m+2}-2$, by defining 
\[
d_{m+1} = \left\{
\begin{array}{cl}
0 & ( 0 \leq i \leq 2^{m+1}-2 ) , \\
1 & ( 2^{m+1}-1 \leq i \leq 2^{m+2}-3 ) , \\
2 & ( i = 2^{m+2}-2 ), 
\end{array}
\right. 
\]
the function $C_m^{-1}$ uniquely maps
  $i - d_{m+1} \left( 2^{m+1} - 1 \right) \in \left\{ 0, 1, \ldots, 2^{m+1}-2 \right\} $
  to the string $d_m d_{m-1} \cdots d_1$.
Therefore $C_{m+1}$ is injective.
Conversely, because the language $\calL_{m+1}$ consists of the strings with the length $m+1$ that are accepted by
  the deterministic (except for permitting zero outdegree) finite automaton in figure \ref{fig:FA}, 
  the number of elements in $\calL_{m+1}$ is given by $\# \calL_{m+1} = \ds (1, 1, 1) A^{m+1} (1, 0, 0)^t = 2^{m+2} - 1$, 
  where 
\[
A=\left(
\begin{array}{ccc}
0 & 0 & 0 \\
2 & 2 & 0 \\
1 & 1 & 1
\end{array}
\right) 
\]
  is the adjacency matrix. 
Thus the map $C_{m+1}$ is shown to be bijective.
\qed
\end{Proof}

\subsection{Properties of sequence $b_l$}

\begin{figure}[tbp]
  \begin{center}
    \includegraphics[width=9cm]{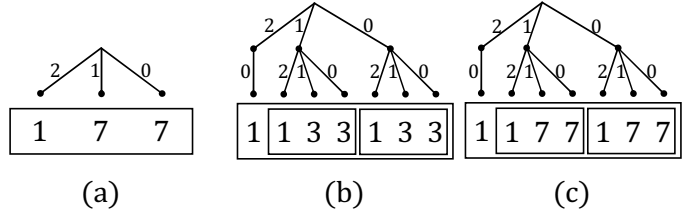} 
    \caption{Examples of tree for (a) $I^{(3)}_{2}$ \  (b) $I^{(3)}_{1}$ \  (c) $I^{(4)}_{2}$. 
      Numbers in lower part correspond to the value of sequence $\left\{ I^{(h)}_{i}(j) \right\}_j$ for each $j$.}
    \label{fig:tree}
  \end{center}
\end{figure}

We prepare some propositions needed to prove theorem \ref{thm:concrete}. 
Let $i$ be $0 \leq i \leq h$. Since the length of $I^{(h)}_{i}$ is $2^{h-i+1}-1$, 
  the $j$-th element $I^{(h)}_{i}(j)$ in the sequence $I^{(h)}_{i}$ from the left
  is located at $(2^{h-i+1}-2-j)$-th position from the right
  ($0 \leq j \leq 2^{h-i+1}-2$, which is numbered from 0).
For this element $I^{(h)}_{i}(j)$, 
  we define the string $\alpha^{(h)}_{i} (j) := C_{h-i}^{-1}(2^{h-i+1}-2-j) \in \calL_{h-i}$.
Suppose a ternary tree whose edge has the label with the alphabet
  $d_k$ at depth $h-i-k$ from root to leaf $I^{(h)}_{i} (j)$, 
  where $d_{h-i} d_{h-i-1} \cdots d_{1} := \alpha^{(h)}_{i} (j) \in \calL_{h-i}$.
In the case $i=h$, the tree consists of only root. 
Examples are shown in table \ref{tbl:I3} and figure \ref{fig:tree}.

\begin{table}[bp]
  \caption{Correspondence among $I^{(h)}_{i}, \alpha^{(h)}_{i}$ and $C_{h-i}(\alpha^{(h)}_{i})$ in the case $h=3$ and $i=1$.
    (cf.~figure \ref{fig:tree}(b))}
  \begin{indented}
  \item[]
%  \begin{tabular}{|c||ccccccc|}
  \begin{tabular}{cccccccc}
\br
$j$ & 0 & 1 & 2 & 3 & 4 & 5 & 6 \\ \mr
$I^{(3)}_{1} (j)$ & 1 & 1 & 3 & 3 & 1 & 3 & 3 \\ 
$\alpha^{(3)}_{1} (j)$ & {\tt 20} & {\tt 12} & {\tt 11} & {\tt 10} & {\tt 02} & {\tt 01} & {\tt 00} \\ 
$C_2(\alpha^{(3)}_{1} (j))$ & 6 & 5 & 4 & 3 & 2 & 1 & 0 \\ \br
  \end{tabular}
  \label{tbl:I3}
  \end{indented}
\end{table}

\begin{Proposition}
Let $i$ and $j$ be the above integers ($0 \leq i \leq h$, $0 \leq j \leq 2^{h-i+1}-2$). Then, 
\begin{equation}
I^{(h)}_{i}(j) = \left\{
  \begin{array}{ll}
    2^{i+1}-1 & ( \alpha^{(h)}_{i} (j) \mbox{ does not contain {\tt 2}} ) \\
    1 & ( \alpha^{(h)}_{i} (j) \mbox{ contains {\tt 2}} )
  \end{array} 
\right. ,
  \label{eq:Ihij}
\end{equation}
\begin{equation}
\sum^{h-i}_{k=1} d_k (2^k - 1) = C_{h-i} \left( \alpha^{(h)}_{i} (j) \right) = 2^{h-i+1}-2-j.
\label{eq:sumdk}
\end{equation}
For the case $i=h$, empty sum is defined to be zero convention, which 
  results in  
  $j=0$, $\alpha^{(h)}_{i}(j) = \varepsilon$, $C_{h-i}(\varepsilon) = 0$, and $I^{(h)}_{i}(j) = 2^{h+1}-1=K$.
\label{prop:correspondence}
\end{Proposition}

\begin{Proof}
Since the map $C_m$ is bijective and preserves the magnitude relation of integers and lexicographic order of strings, 
  there is one-to-one correspondence between $\alpha^{(h)}_{i} (j)$ and $I^{(h)}_{i} (j)$ for each $i$.
The property that $\calL$ have, 
  "if there exists $k$ such that $d_{k}={\tt 2}$, then $d_{k-1} = d_{k-2}= \cdots = d_{1} = {\tt 0}$",
  implies that, if {\tt 2} appears in $\alpha^{(h)}_{i} (j)$ then only {\tt 0}'s may follow after the {\tt 2}.
In the definition of $I^{(h)}_{i}$, this restriction corresponds to the rule that the integer 1's are never replaced.
On the other hand, the integers that may be replaced in the definition are $(2^{i+1} - 1)$'s. 
(\ref{eq:sumdk}) follows from the proposition \ref{prop:bijection}. 
\qed
\end{Proof}
Note that in the case $i=0$, $I^{(h)}_{i}(j)=1$ holds for all $j$, 
  regardless of the characters in the string $\alpha^{(h)}_{i} (j)$ because of  $2^{i+1} - 1 = 1$.

\begin{Lemma}
Suppose the string
$d_{h-i} d_{h-i-1} \cdots d_{1} := \alpha^{(h)}_{i} (j) \in \calL_{h-i}$ 
with  $0 \leq i < h$, $0\leq j < h-i $. 
Then, $j-\sum_{k=1}^{h-i} d_k = -2$ and $I^{(h)}_{i}(j) = 1$.
\label{lem:firstn}
\end{Lemma}

\begin{Proof}
Each $\alpha^{(h)}_{i} (j)$ is as follows: 
$\alpha^{(h)}_{i} (0) = {\tt 20}^{h-i-1} = {\tt 2}\overbrace{{\tt 0}\cdots {\tt 0}}^{h-i-1}$,
  $\alpha^{(h)}_{i} (1) = {\tt 120}^{h-i-2}$,
  $\ldots, \alpha^{(h)}_{i} (j) = {\tt 1}^j {\tt 20}^{h-i-j-1}, \ldots,
  \alpha^{(h)}_{i} (h-i-1) = {\tt 1}^{h-i-1}{\tt 2}$.
The number of the characters {\tt 2} in $\alpha^{(h)}_{i} (j)$ is always $1$ and that of {\tt 1} is $j$.
\qed
\end{Proof}

\begin{Lemma}
Given $m \in \Z_{\geq 1}$,
  and suppose that the strings $d_m d_{m-1} \cdots d_1$ and $d'_m d'_{m-1} \cdots d'_1 \in \calL_m$ satisfy
  $C_m(d_m d_{m-1} \cdots d_1)+C_m(d'_m d'_{m-1} \cdots d'_1) = 2^{m+1}-m-1$.
Then, $\sum_{k=1}^{m}(d_k + d'_k) = m+1$. 
\label{lem:pairs}
\end{Lemma}
\begin{Proof}
Let $P(m)$ be this lemma \ref{lem:pairs} with $m$ that we prove by induction.
The assertion is clearly true for $m=1$ since $(d_1, d'_1) = (0, 2), (1, 1), (2, 0)$. Therefore $P(1)$ is correct. 
Assume that $P(m)$ is correct and suppose
$C_{m+1}(d_{m+1} d_m \cdots d_1)+C_{m+1}(d'_{m+1} d'_m \cdots d'_1) = 2^{m+2}-m-2$.
Because 
$C_{m+1}(d_{m+1} d_m \cdots d_1) = d_{m+1}(2^{m+1}-1) + C_m(d_{m} \cdots d_1)$ and 
$C_{m+1}(d'_{m+1} d'_m \cdots d'_1) = d'_{m+1}(2^{m+1}-1) + C_m(d'_{m} \cdots d'_1)$, 
we obtain
$( d_{m+1}+d'_{m+1} ) (2^{m+1}-1) + C_m(d_{m} \cdots d_1) + C_m(d'_{m} \cdots d'_1) = 2^{m+2}-m-2$. 
\begin{itemize}
\item If $ d_{m+1}+d'_{m+1} = 2 $, 
  $C_m(d_{m} \cdots d_1) + C_m(d'_{m} \cdots d'_1) = -m < 0$ does not satisfy the condition.
\item If $ d_{m+1}+d'_{m+1} = 1 $,
  $C_m(d_{m} \cdots d_1) + C_m(d'_{m} \cdots d'_1) = 2^{m+1}-m-1$ holds. Then by $P(m)$, 
  $\sum_{k=1}^{m}(d_k + d'_k) = m+1$. Therefore
  $\sum_{k=1}^{m+1}(d_k + d'_k) = m+2$．
\item If $ d_{m+1}+d'_{m+1} = 0 $, 
  $C_m(d_{m} \cdots d_1) + C_m(d'_{m} \cdots d'_1) = 2^{m+2}-m-2$ holds. 
Because $C_m(d_{m} \cdots d_1)$ and $C_m(d'_{m} \cdots d'_1)$ are bounded
  as $C_m(d_{m} \cdots d_1),  C_m(d'_{m} \cdots d'_1) \leq 2^{m+1}-2$ by definition, 
  these are also bounded from below; $2^{m+1}-m \leq C_m(d_{m} \cdots d_1),  C_m(d'_{m} \cdots d'_1)$.
Introducing $j := 2^{m+1}-2-C_m(d_{m} \cdots d_1)$ and $j' := 2^{m+1}-2-C_m(d'_{m} \cdots d'_1)$ 
  yields $\sum_{k=1}^{m} d_k = j + 2$ and $\sum_{k=1}^{m} d'_k = j' + 2$ from lemma \ref{lem:firstn}.
Since $\sum_{k=1}^{m}(d_k + d'_k) = j+j'+4 = m+2$ and $d_{m+1}+d'_{m+1} = 0$, we obtain
  $\sum_{k=1}^{m+1}(d_k + d'_k) = m+2$．
\end{itemize}
Hence $P(m+1)$ holds, and this ends the proof by induction.
\qed
\end{Proof}

\begin{Proposition}
Given $1 \leq i \leq h$ and $0\leq j \leq 2^{h-i+1}-2 $, and suppose
  the strings $\beta := \alpha^{(h)}_{i} (j) \in \calL_{h-i}$
  and $\beta' := \beta {\tt 0} \in \calL_{h-i+1}$, i.e., the concatenation of $\beta$ and ${\tt 0}$. 
Then, 
\begin{equation}
\ds
\sum_{k=0}^{j} I^{(h)}_{i}(k) = 
  \sum_{k'=0}^{j'} I^{(h)}_{i-1} (k'), 
\label{eq:jjsame}
\end{equation}
where $j' := 2^{h-i+2}-2-C_{h-i+1}(\beta')$.
  \label{prop:sum1}
\end{Proposition}
\begin{Proof}
Now we are considering the case $i \geq 1$, thereby $2^{i+1}-1 \neq 1$ holds.
Therefore $I^{(h)}_{i}(j) = 2^{i+1}-1$ if and only if 
$\beta = \alpha^{(h)}_{i} (j) $ does not contain {\tt 2} 
from (\ref{eq:Ihij}).

Let us recall that when the sequence $I^{(h)}_{i-1}$ is constructed, 
  each value $2^{i+1}-1$ in $I^{(h)}_{i}(j)$ is replaced by \underline{$(1, 2^i-1, 2^i-1)$}. 
In other words, when $\beta$ does not contain {\tt 2}, 
  the three elements specified by $j'$ in the sequence $I^{(h)}_{i-1}$ are determined as follows:
\begin{equation*}
\left\{
\begin{array}{rcl}
I^{(h)}_{i-1} (j'-2) & = & 1, \\
I^{(h)}_{i-1} (j'-1) & = & 2^i-1, \\
I^{(h)}_{i-1} (j') & = & 2^i-1, 
\end{array}
\right.
\end{equation*}
where the corresponding strings are
\begin{equation*}
\left\{
\begin{array}{rcl}
  \alpha^{(h)}_{i-1} (j'-2) & = & \beta {\tt 2}, \\
  \alpha^{(h)}_{i-1} (j'-1) & = & \beta {\tt 1}, \\
  \alpha^{(h)}_{i-1} (j') & = & \beta {\tt 0}, 
\end{array}
\right.
\end{equation*}
respectively. Therefore if $\beta$ does not contain {\tt 2}, then 
\begin{equation}
I^{(h)}_{i} (j) = \sum_{k'=j'-2}^{j'} I^{(h)}_{i-1} (k') = 2^{i+1}-1.
\label{eq:c01}
\end{equation}
Next, consider the case $I^{(h)}_{i}(j) = 1$. Since $\beta$ contains {\tt 2},
this value $1$ is not replaced as it is:
\begin{equation}
  I^{(h)}_{i} (j) = I^{(h)}_{i-1} (j') = 1.
\label{eq:c2}
\end{equation}
Note that $\beta {\tt 1}, \beta {\tt 2} \not\in \calL_{h-i+1}$ in this case.

In the left-hand side of (\ref{eq:jjsame}), the equality (\ref{eq:c01}) or (\ref{eq:c2}) holds,
  depending on whether $\alpha^{(h)}_{i}(k)$ contain {\tt 2} ($0 \leq k \leq j$).
Conversely in the right-hand side of (\ref{eq:jjsame}),
  suppose the string $\gamma := \alpha^{(h)}_{i-1}(k') = C_{h-i+1}^{-1}(2^{h-i+2}-2-k') \in \calL_{h-i+1}$
  for $0 \leq k' \leq j'$.
Since $h-i+1 \geq 1$, 
  the string $\gamma $ with the length more than zero necessarily have the parent node $\mbox{del}(\gamma)$,
  which is obviously unique. 
Thus the proposition \ref{prop:sum1} holds.
\qed
\end{Proof}

\begin{Corollary}
For $i$, $j$, and $j'$ in proposition \ref{prop:sum1}, if $0 \leq j < 2^{h-i+1} - 2$, then
\[ \ds
  \sum_{k=j+1}^{2^{h-i+1}-2} I^{(h)}_{i}(k) + \sum_{k'=0}^{j'} I^{(h)}_{i-1}(k') = K.
\]
Otherwise $j = 2^{h-i+1} - 2$, $j' = 2^{h-i+2} - 2$, and 
\[ \ds
  \sum_{k'=0}^{2^{h-i+2}-2} I^{(h)}_{i-1}(k') = K.
\]
\label{cor:sum2n}
\end{Corollary}
\begin{Proof}
The equalities $ \ds \sum_{k=0}^{2^{h-i+1}-2} I^{(h)}_{i} (k) = \sum_{k'=0}^{2^{h-i+2}-2} I^{(h)}_{i-1} (k') = K $
hold from (\ref{eq:sum_ini}). Substituting (\ref{eq:jjsame}) into these equations yields this corollary.
\qed
\end{Proof}
This corollary \ref{cor:sum2n} suggests the intervals such that the sum of $b_k$ equals $K$.
The following proposition \ref{prop:ToLeft} provides the left end of the interval 
  as the right end fixed such that the sum of $b_k$ is less than or equal to $K$.
In the opposite direction, the proposition \ref{prop:ToRight} provides the right end as the left end fixed
  such that the sum of $b_k$ is equal to $K$.

\begin{Proposition}
Given $0 \leq i \leq h-1$ and $0 \leq j \leq 2^{h-i+1}-2$, and suppose 
  $d_{h-i} d_{h-i-1} \cdots d_{1} := \alpha^{(h)}_{i} (j) \in \calL_{h-i}$. 
We then obtain 
  $\ds \mbox{del} \left(\alpha^{(h)}_{i} (j) \right) = \alpha^{(h)}_{i+1} (j') \in \calL_{h-i-1}$, 
  where $\ds j' := \frac{1}{2} \left( j + \sum_{k=1}^{h-i} d_k \right)  - 1$ and 
  $\ds \left( j + \sum_{k=1}^{h-i} d_k \right)$ is always even integer.
This yields 
$ \ds \sum_{k = j_1+1}^{j_2 + d_1} b_k = K$ and $\ds \sum_{k = j_1}^{j_2} b_k > K$,
  where
  $j_1 := j' + 2^{h-i}-(h-i+1)$, 
  $j_2 := j + 2^{h-i+1}-(h-i+2)$.
The interval is given as $j_2-j_1 = \ds \frac{1}{2} \left( j - \sum_{k=1}^{h-i} d_k \right) +2^{h-i} $.
\label{prop:ToLeft}
\end{Proposition}
\begin{Proof}
Since $\ds (2^{h-i+1}-2) - j = \sum_{k=1}^{h-i} d_k (2^k -1)$ from (\ref{eq:sumdk}),
we may find 
$\ds \left( j + \sum_{k=1}^{h-i} d_k \right)$ is even.
Substituting 
$\ds j' = \frac{1}{2} \left( j + \sum_{k=1}^{h-i} d_k \right)  - 1$ into the above equation yields
$\ds (2^{h-i}-2) - j' = \sum_{k=1}^{h-i-1} d_{k+1} (2^k -1) $. 
The right-hand side of this equation is equal to $C_{h-i-1} ( \mbox{del} (\alpha^{(h)}_{i} (j)))$, while 
 the left-hand side $C_{h-i-1} ( \alpha^{(h)}_{i+1} (j') )$. Therefore we obtain
$\mbox{del} (\alpha^{(h)}_{i} (j)) = \alpha^{(h)}_{i+1} (j')$.

Substituting $j_1$ for $j$ in the corollary \ref{cor:sum2n} and rewriting by $b_k$,
  we obtain $\sum_{k = j_1+1}^{j_2 + d_1} b_k = K$.
Both ends $j_1$ and $j_2$ are determined by the relations 
  $I^{(h)}_{i+1}(j') = b_{j_1}$ and $I^{(h)}_{i}(j) = b_{j_2}$, which follow from (\ref{eq:defb}).

The inequality in this proposition is obtained as follows: if $d_1={\tt 0}$, then $\sum_{k = j_1}^{j_2} b_k > K$ because $b_{j_1} > 0$.
If $d_1={\tt 1}$ or ${\tt 2}$, then 
  $b_{j_1} = 2^{i+2}-1 > \sum_{k'=1}^{d_1} b_{j_2+k'}$ and therefore $\sum_{k = j_1}^{j_2} b_k > K$,
  because $\mbox{del} (\alpha^{(h)}_{i} (j)) = \alpha^{(h)}_{i+1} (j')$ does not contain {\tt 2}. 
\qed
\end{Proof}

\begin{Proposition}
Given $1 \leq i \leq h$ and $0 \leq j \leq 2^{h-i+1}-2$, and suppose
  $\beta := d_{h-i} d_{h-i-1} \cdots d_{1} := \alpha^{(h)}_{i} (j) \in \calL_{h-i}$. 
We then obtain $\beta {\tt 0} = \alpha^{(h)}_{i-1} ( j') \in \calL_{h-i+1}$, 
  where $\ds j' = 2j+2 - \sum_{k=1}^{h-i} d_k$. 
This yields $\ds \sum_{k = j_1+1}^{j_2} b_k = K$, 
where 
  $j_1 := j + 2^{h-i+1} - (h-i+2) $,
  $j_2 := j' + 2^{h-i+2} - (h-i+3) $. 
The interval is given as $\ds j_2-j_1 = j+1+2^{h-i+1} - \sum_{k=1}^{h-i} d_k$.
\label{prop:ToRight}
\end{Proposition}
\begin{Proof}
The proof is similar to the one in proposition \ref{prop:ToLeft}. 
Substituting $d_1={\tt 0}$ in proposition \ref{prop:ToLeft} and solving it in $j'$ instead of $j$,
  the first half  of this proposition is obtained.
In this case, though $\beta {\tt 0} \in \calL_{h-i+1}$ always hold, 
  $\beta {\tt 1}, \beta {\tt 2} \in \calL_{h-i+1}$ is not always true because $\beta$ may contain {\tt 2}.
The latter half of this proposition follows from rewriting corollary \ref{cor:sum2n} by $b_k$
  via the correspondence (\ref{eq:defb}).
\qed
\end{Proof}

\subsection{Proof of theorem \ref{thm:concrete}}
In this subsection, we prove theorem \ref{thm:concrete}. 
Substituting $\Om = K+1 = 2^{h+1}$ into (\ref{eq:travelS}) yields
\begin{equation}
  S(\xi) - S(\xi-1) = M \left( 0, S(\xi+K)-S(\xi-K-1)-1 \right).
  \label{eq:ss}
\end{equation}
Note that this equation is quadratic and does not define explicit evolutions. 
Using $S(\xi) = \# \left\{ \mu \in J \left| \ \mu \leq \xi \right. \right\} \bmod 3$ 
  with the set $J$, we obtain
\[ \ds
  S(\xi )-S(\xi-1) =   \left\{
  \begin{array}{ll}
    1 & ( \xi \in J) \\
    0 & \mbox{otherwise}
  \end{array} , 
  \right.
\]
\[
S(\xi +K)-S(\xi-K-1) = W(\xi) \bmod 3,
\]
where $W(\xi) := \# \left\{ \mu \in J \left| \ \xi-K \leq \mu \leq \xi+K \right. \right\}$. 
Thus the problem is reduced to finding the number of elements of $J$ in the width $2K+1$ window.
With $a_0=0, a_1=K$ in mind, we obtain 
$W(\xi) = 0$ for $\xi < -K$, and $W(\xi) = 1$ for $-K \leq \xi < 0$.
Therefore (\ref{eq:ss}) holds for all $\xi < 0$.
Next, for $0 \leq \xi < a_r$, the integer $\xi$ may be uniquely expressed by the pair $(l, \eta)$ 
  as $\xi = a_l + \eta$ ($0 \leq l < r$, $0\leq \eta < b_l$) 
  since $a_l= \sum_{k=0}^{l-1} b_{k}$ and $a_0=0$.
Moreover because the integer sequence $\ds \left\{ b_l \right\}_{l=0}^{r-1}$ is palindromic, 
  $0 \leq l \leq \ds \left\lceil \frac{r}{2} \right\rceil$ is sufficient for $\xi$.
Let us denote $W(l, \eta) := W(\xi = a_l + \eta) $. By using
\[
a_m - a_l = \left\{
\begin{array}{ll}
\sum_{k=l}^{m-1} b_k & (l < m) \\
 0 & (l = m), \\
-\sum_{k=m}^{l-1} b_k & (l > m)
\end{array}
\right.
\]
we obtain
\begin{eqnarray}
W(l, \eta)  
  & = & \# \left\{ \mu \in J \left| \ a_l + \eta-K \leq \mu \leq a_l + \eta+K \right. \right\} \nonumber \\
  & = & \# \left\{ m \left| \ 0 \leq m \leq r, a_l + \eta-K \leq a_m \leq a_l + \eta+K \right. \right\} \nonumber\\
  & = & \# \left\{ m \left| \ 0 \leq m \leq r, (a_m - a_l )-K \leq \eta \leq (a_m - a_l)+K \right. \right\} \nonumber \\
  & = & 1 + \# \left\{ m \left| \ 0 \leq m < l \leq r, \eta \leq K - \sum_{k=m}^{l-1} b_k \right. \right\} \nonumber \\
  &   & \hspace*{3mm} + \# \left\{ m \left| \ 0\leq l < m \leq r, \sum_{k=l}^{m-1} b_k \leq K+\eta \right. \right\},
    \label{eq:Wle}
\end{eqnarray}
where we have used $0 \leq \eta < b_l \leq K$.
In \ref{sec:eta0} and \ref{sec:etaneq0}, 
\begin{equation}
  W(l, \eta) \equiv \left\{
  \begin{array}{ll}
    2 & (\eta = 0) \\
    0, 1 & (1 \leq \eta < b_l )
  \end{array} \right.
 \pmod 3
 \label{eq:Wval}
\end{equation}
is shown for $0 \leq l \leq \ds \left\lceil \frac{r}{2} \right\rceil$.
Thus (\ref{eq:ss}) follows from the fact \ref{fact:M},
  and this ends the proof of theorem \ref{thm:concrete}.
\qed

\subsection{Patterns at a certain time}
\label{sec:patterns}
In the previous section, we have numerically observed the whole patterns of the travelling waves
  in figures \ref{fig:ffBBS0102} and \ref{fig:ffBBS03}.
In this subsection, we show that the one-soliton solution with fractional velocity 
  mentioned in theorem \ref{thm:concrete} contains the pattern `$12^h0^h1$'  at a certain time. 
For any positive integer $h$, the travelling wave with the dispersion relation $\Om = K+1 = 2^{h+1}$ 
  always exists and is given by
  $S(\xi) = \# \left\{ \mu \in J \left| \ \mu \leq \xi \right. \right\} \bmod 3$.
Define $V(\xi) := \# \left\{ \mu \in J \left| \ \xi-K+1 \leq \mu \leq \xi \right. \right\} \in \Z$.
Then the dependent variables in ffBBS (\ref{eq:ffBBS}) with $p=3$ is given by 
  $U^{t}_{n}=V(\xi) \bmod 3$, where $\xi = K n - \Om t$.
Let us fix the time $t$ and observe $V(\xi)$ with respect to the space coordinate $n$.
For an appropriate $t$, we may choose $\xi = K n + (K-1)/2$ and obtain the following:
\[
V \left( K n + \frac{K-1}{2} \right) = \left\{
\begin{array}{ll}
0 & (n < 0) \\
1 & (n = 0) \\
3 \cdot 2^{n-1}-1 & (1 \leq n \leq h) \\
3 \cdot 2^{2h-n} & (h+1 \leq n \leq 2 h) \\
1 & (n = 2h+1) \\
0 & (2h+1 < n) 
\end{array}
\right. .
\]
Thus for any $h$, the dependent variables $\left\{ U^t_n \right\}_n$ have the following pattern by taking modulo 3.
\[
\left\{ U^t_n \right\}_{n \in \Z} =
  \left\{ \ldots, 0, 0, 1, \overbrace{2, \ldots, 2}^{h}, \overbrace{0, \ldots, 0}^{h}, 1, 0, 0, \ldots \right\}
\]
Figure \ref{fig:ffBBS03} is numerically calculated by these patterns with $h=1$, $2$, and $3$ as initial ones.

\begin{Proof}
First, keeping $h \geq 1$ in mind, 
  we obtain $V(\xi) = 0$  for $\xi < 0$, $V((K-1)/2) = \# \left\{a_0 = 0\right\} = 1$, and
  $V(K+(K-1)/2) = \# \left\{a_1 = 2^{h+1}-1, a_2 = 2^{h+1} \right\} = 2$, 
  since $b_0 = 2^{h+1}-1, b_1 = 1, b_2=2^{h}-1, b_3=2^{h}-1$. 

Next, we examine the case $2\leq n \leq h$. Let us define $i:=h-n$, then $0 \leq i \leq h-2$.
Proposition \ref{prop:correspondence} yields $\beta = \alpha^{(h)}_{i} (j)$, where 
  $\beta := {\tt 1}{\tt 0}^{h-i-1} \in \calL_{h-i}$ and $j := 2^{h-i+1}-2-C_{h-i}(\beta) = 2^{h-i}-1$.
By repetition of proposition \ref{prop:sum1}, we may delete the {\tt 0}'s from the tail of $\beta$ and obtain
\[
  \sum_{k=0}^{j} I^{(h)}_{i} (k) = \cdots = \sum_{k=0}^{1} I^{(h)}_{h-1} (k) = 1+(2^h-1) = 2^h = \frac{K+1}{2}.
\]
The correspondence $b_l = I^{(h)}_{i}(j)$ follows from (\ref{eq:defb}), where
  $l := j+2^{h-i+1}-(h-i+2) = 3\cdot 2^{h-i}-(h-i)-3$. 
From (\ref{eq:sum_ini}), we therefore obtain 
\[
  K(h-i)+\frac{K+1}{2} = 
  \sum_{m=i+1}^{h} \sum_{k=0}^{2^{h-m+1}-2} I^{(h)}_{m} (k) + \sum_{k=0}^{j} I^{(h)}_{i} (k) = \sum_{k=0}^{l} b_k = a_{l+1}, 
\]
and 
\begin{eqnarray*}
  V(\xi) & = & \# \left\{ \mu \in J \left| \ \xi-K+1 \leq \mu \leq \xi \right. \right\} \\
    & = & \# \left\{ m \left| \ 0\leq m \leq r, \xi-K+1 \leq a_m \leq \xi \right. \right\} \\
    & = & \# \left\{ m \left| \ 0\leq m \leq r, K(n-1)+\frac{K+1}{2} \leq a_m < Kn+\frac{K+1}{2} \right. \right\} \\
    & = & \# \left\{ m \left| \ 3\cdot 2^{n-1}-n-1 \leq m < 3\cdot 2^{n}-n-2 \right. \right\} \\
    & = & 3\cdot 2^{n-1}-1, 
\end{eqnarray*}
where $\xi = K n + (K-1)/2$ ($2\leq n \leq h$). 
Similarly for $n > h$, we may calculate $V(\xi)$ by means of palindromic property of $\left\{ b_l \right\}$. 
\qed
\end{Proof}

\section{Concluding remarks}
\label{sec:remarks}

In this paper, we have proposed the solitonic systems over finite fields (ffBBS)
  with respect to an analogue of bilinear form of BBS.
We have also constructed the one-soliton solutions of (\ref{eq:ffBBS}),
  which is categorized as the context-free language, 
  since $I^{(h)}$ is palindromically defined in (\ref{eq:Iseq}) for any positive integer $h$.
As with the conserved quantities of BBS described by Dyck language \cite{TTS, T91}, 
  this fact may relate with the integrability of our systems.

For the periodic BBS (pBBS; \cite{YT02}), the asymptotic behaviour of fundamental cycle of pBBS
  was investigated\cite{MT03}.
In their study, the order of the maximum cycle is $\exp\sqrt{K}$ with respect to the system size $K$. 
Furthermore almost all initial states have the fundamental cycle less than $\exp [(\log K)^2]$.
These cycles are extremely short compared with the number of states $2^K$ and 
  considered to be the consequence of the integrability of the pBBS.
On the other hand, one-soliton solutions in theorem \ref{thm:concrete} have the period
  at most $\sim 2^{K/2}$. From this fact, indeed our ffBBS proposed in this paper is solitonic system, 
  though it might be non-integrable. 

The numerical experiments nonetheless show the preserving of the pattern before and 
  after the collisions in figures \ref{fig:ffBBS0102} and \ref{fig:ffBBS03}. 
The ffBBS should therefore have at least some conserved quantities in order to preserve the solitary patterns.
Since theorem \ref{thm:concrete} states that even one-soliton solutions have 
  quite a complex structure with nested fractals,
  the elucidation of the solutions in ffBBS with respect to its conserved quantities and periods 
  may lead to new discoveries about a concept and mechanism of integrability over ultradiscretization or finite fields.

The key formula for ultradiscretization (or tropical) method is the limiting procedure\cite{TTMS}:
\[
  \max(X, Y) = \lim_{\epsilon \to +0} \epsilon \log \left( e^{X/\epsilon} + e^{Y/\epsilon} \right), 
\]
where $X, Y \in \R$. 
A naive analogue of this procedure for finite fields may be 
  $f(x, y) = \log_g \left( g^{x} + g^{y} \right) $, 
  where $g$ is an element of $\F_q$ and $\log_g$ the discrete logarithm.
Such a trial, however, does not apparently go well because $g^{x} + g^{y}$ may be zero.
Our method proposed in this paper is widely applicable to the existing systems, 
  which is not limited to solitonic systems. 
At least the systems that are written by max-plus algebra may be translated to the ones
  over finite fields with a function $M$.
In general, such translated systems may not be necessarily meaningful.
Though, as far as we showed in this paper, a novel system is systematically obtained,
  which have soliton solutions with fractal structures. 
More trials for general systems will shed light on the systems over finite fields
  such as pseudo-random number generators, coding theory and cryptography.

\ack
This work was supported by JSPS KAKENHI Grant Number 19740053 and 23611027. 
The author would like to thank Hideyuki Nakashima for the above grant 23611027. 

\appendix

\section{Calculations of (\ref{eq:Wval})}

Since the calculations of (\ref{eq:Wval}) are straight but lengthy, we show the 
  details in appendix.
In the following subsections, the cases $\eta = 0$ and $\eta \neq 0$
  are examined respectively.

\subsection{The case $\eta = 0$:}
\label{sec:eta0}
In this case, (\ref{eq:Wle}) turns into
\begin{eqnarray*}
W(l, 0) 
   & = & 1 + \# \left\{ m \left| 0 \leq m < l \leq r, \sum_{k=m}^{l-1} b_k  \leq K \right. \right\} \\
   & & \ + \# \left\{ m \left| 0\leq l < m \leq r, \sum_{k=l}^{m-1} b_k \leq K \right. \right\} .
\end{eqnarray*}
First, we separately calculate for the small $l$;
  $W(0, 0) = 2$ for $l=0$, and $W(1, 0) = 5$ for $l=1$,
  which corresponds to $W(\xi=a_1=15)=5$ in example \ref{ex:n3}.

Next, for 
$2 \leq l \leq 2^{h+1}-(h+2)$, 
  we have $b_{l-1} = I^{(h)}_{i} (c)$ with $i$ such that $1 \leq i \leq h-1$ since  
  $2^{h-i+1}-(h-i+2) \leq l-1 < 2^{h-i+2}-(h-i+3)$, 
  where $c := (l-1)-2^{h-i+1}+(h-i+2)$. 
In this case, both propositions \ref{prop:ToLeft} and \ref{prop:ToRight}
  are applicable due to the condition $1 \leq i \leq h-1$.
Then, by substituting $c$ for $j$ in proposition \ref{prop:ToLeft} and \ref{prop:ToRight},
  we obtain
\begin{eqnarray*}
  W(l, 0) & = & 1 + \left\{ \frac{1}{2} \left( c-\sum_{k=1}^{h-i} d_k \right) + 2^{h-i} \right\}
    + \left\{ c+1+2^{h-i+1}-\sum_{k=1}^{h-i} d_k \right\} \\
  & = & 3 \left( 2^{h-i+1} - \sum_{k=1}^{h-i} d_{k} 2^{k-1} \right) -1, 
\end{eqnarray*}
and therefore $W(l, 0) \equiv 2 \bmod 3$.

Next, for $ 2^{h+1}-(h+1) \leq l \leq 2^{h+1}-2 $, we obtain 
\begin{equation*}
\# \left\{ m \left| \ 0\leq l < m \leq r, \sum_{k=l}^{m-1} b_k \leq K \right. \right\} = 2^{h+1} -1 = K, 
\end{equation*}
  since $b_{l} = b_{l+1} = \cdots = b_{l+2^{h+1}-2} = 1$. 
Defining $c := (l-1)-2^{h+1}+(h+2)$ and substituting $c$ for $j$ in proposition \ref{prop:ToLeft}
  with $i=0$ leads to
\begin{equation*}
 \# \left\{ m \left| \ 0 \leq m < l \leq r, \sum_{k=m}^{l-1} b_k  \leq K \right. \right\}
  =  \frac{1}{2} \left( c - \sum_{k=1}^{h} d_k \right) + 2^{h}.
\end{equation*}
From lemma \ref{lem:firstn} due to $0 \leq c < h$, we obtain
\begin{eqnarray*}
  W(l, 0) & = & 1 + \frac{1}{2} \left( c - \sum_{k=1}^{h} d_k \right) + 2^{h}
    + (2^{h+1}-1) \\
  & = & 3 \cdot 2^h - 1. 
\end{eqnarray*}

Finally, we calculate for the case $ 2^{h+1}-1 \leq l \leq \ds \left\lceil \frac{r}{2} \right\rceil $. 
Because some $b_k$'s in the third term of the right-hand side of (\ref{eq:Wle}) are located 
  in $I^{(h)R}_1$, we rewrite by means of the symmetry $b_k = b_{r-k-1}$ as 
\begin{eqnarray*}
   \# \left\{ m \left| 0\leq l < m \leq r, \sum_{k=l}^{m-1} b_k \leq K \right. \right\} \\
   \hspace*{10mm} = \# \left\{ m \left| 0\leq r-m < r-l \leq r, \sum_{k=r-m}^{r-l-1} b_k \leq K \right. \right\}, 
\end{eqnarray*}
and $b_{r-l-1} = I^{(h)}_0 (c)$, where $c := (r-l-1)-2^{h+1}+(h+2)$.
Substituting $c$ for $j$ in proposition \ref{prop:ToLeft} leads to 
\begin{eqnarray*}
   \# \left\{ m \left| \ 0\leq l < m \leq r, \sum_{k=l}^{m-1} b_k \leq K \right. \right\} & = &
   \ds \frac{1}{2} \left(c - \sum_{k=1}^{h} d_k \right) + 2^h, 
\end{eqnarray*}
where $d_{h} \cdots d_{1} := \alpha^{(h)}_{0} (c)$ and $\sum_{k=1}^{h} d_k (2^k-1) = 2^{h+1}-2-c$. 
On the other hand, for the second term of the right-hand side of (\ref{eq:Wle}), 
  since $b_{l-1} = I^{(h)}_0 (c')$ and $\sum_{k=1}^{h} d'_k (2^k-1) = 2^{h+1}-2-c'$,  
  where $c' := (l-1)-2^{h+1}+(h+2)$ and $d'_{h} \cdots d'_{1} := \alpha^{(h)}_{0} (c')$, 
  substituting $c'$ for $j$ in proposition \ref{prop:ToLeft} leads to 
\begin{eqnarray*}
   \# \left\{ m \left| \ 0 \leq m < l \leq r, \sum_{k=m}^{l-1} b_k  \leq K \right. \right\}
& = &
   \ds \frac{1}{2} \left(c' - \sum_{k=1}^{h} d'_k \right) + 2^h.
\end{eqnarray*}
We thus obtain
\begin{eqnarray*}
  W(l, 0) & = & 1 + 
    \left\{ \frac{1}{2} \left(c - \sum_{k=1}^{h} d_k \right) + 2^h \right\} +
    \left\{ \frac{1}{2} \left(c' - \sum_{k=1}^{h} d'_k \right) + 2^h \right\} \\
  & = & 
    3\cdot 2^h + \frac{1}{2} \left\{ h-1-\sum_{k=1}^{h} (d_k + d'_k)  \right\} \\
  & = & 3 \cdot 2^h -1, 
\end{eqnarray*}
  where the last equality follows from lemma \ref{lem:pairs} 
  by making use of $\sum_{k=1}^{h} (d_k+d'_k) (2^k-1) = 2^{h+1}-h-1$. 

All cases in this subsection are summarized as $W(l, 0) \equiv 2 \bmod 3$, 
  which is (\ref{eq:Wval}) for $\eta = 0$. 

\subsection{The case $\eta \neq 0$:}
\label{sec:etaneq0}
In this subsection, we examine the case $\eta \neq 0$ in (\ref{eq:Wle}). 
Since $\xi = a_l + \eta$ and $0\leq \eta < b_l$,
  it is sufficient that we assume $b_l > 1$ for $l$ in $W(l, \eta)$.
That is, we may limit ourselves to $0 \leq l \leq 2^{h+1}-(h+3) $, 
  which have a possibility of $b_l > 1$ in the first half of the sequence $I^{(h)}$. 
In this case, $1 \leq \exists i \leq h$ such that $2^{h-i+1}-(h-i+2) \leq l < 2^{h-i+2}-(h-i+3) $.
For this $i$, let us define $c := l-2^{h-i+1}+(h-i+2)$.
Then $b_{l} = I^{(h)}_i (c) = 2^{i+1}-1$ holds,
  where the last equality is from (\ref{eq:Ihij}) since $b_{l} > 1$.
Therefore it is sufficient that we concentrate on the case 
  the sequence $\alpha^{(h)}_{i} (c) = C_{h-i}^{-1} ( 2^{h-i+1} - 2 -c ) \in \calL_{h-i}$
  does not contain the character {\tt 2}.

First, in the case $i=h$, this condition yields $l=0$.
  Therefore the range of $\eta$ is specified as $1 \leq \eta < b_l = K$.
With $b_0=K=2^{h+1}-1$, $b_1=1$, and $b_2=b_3=2^{h}-1$ in mind, we obtain 
\begin{eqnarray*}
W(l = 0, \eta) 
  & = & 1 + \# \left\{ m \left| \ 0 < m \leq r, \sum_{k=0}^{m-1} b_k \leq K+\eta \right. \right\} \\
  & = & 2 + \# \left\{ m \left| \ 1 < m \leq r, \sum_{k=1}^{m-1} b_k \leq \eta \right. \right\} \\
  & = & 2 + \theta (\eta -1) + \theta (\eta-2^{h} )
\end{eqnarray*}
from (\ref{eq:Wle}).
This implies $W(0, \eta) \not\equiv 2 \bmod 3$ for $1 \leq \eta < b_0 = K$. 

Next, let us examine for $1 \leq i \leq h-1$.
In this case, notice that propositions \ref{prop:ToLeft} and \ref{prop:ToRight} are applicable.
Because all sequences $\alpha^{(h)}_{i} (c) \in \calL_{h-i}$ are longer than one, 
  we may define $d_{h-i} d_{h-i-1} \cdots d_{1} := \alpha^{(h)}_{i} (c)$ and 
  $\beta := d_{h-i} \cdots d_{2} = \mbox{del} (\alpha^{(h)}_{i} (c))$. 
For the second term of the right-hand side of (\ref{eq:Wle})
\begin{equation}
  \# \left\{ m \left| \ 0 \leq m < l \leq r, \eta + \sum_{k=m}^{l-1} b_k \leq K \right. \right\},  
  \label{eq:WleSecond}
\end{equation}
the following properties hold;
Since $d_1 \neq {\tt 2}$, 
\begin{itemize}
\item In the case $d_1 = {\tt 0}$, $\sum_{k=j_1+1}^{l} b_k = K$ follows from proposition \ref{prop:ToLeft}, 
  where $j_1$ is determined by $l$ in the proposition and gives the lower bound of $k$. 
The inequality $\eta < b_l$ yields 
  $ \eta + \sum_{k=j_1+1}^{l-1} b_k < b_l + \sum_{k=j_1+1}^{l-1} b_k = K$.
\item In the case $d_1 = {\tt 1}$, since $b_l$ corresponds to $\beta {\tt 1} ( = \alpha^{(h)}_{i} (c) )$,
$b_{l+1}$ corresponds to $\beta {\tt 0}$. 
Because the strings $\beta {\tt 0}$ and $\beta {\tt 1}$ does not contain the character {\tt 2},
  this leads to $b_{l}=b_{l+1} = 2^{i+1}-1$. 
From proposition \ref{prop:ToLeft}, we then obtain
  $\sum_{k=j_1+1}^{l+1} b_k = K$. ($j_1$ is determined by $l$ in the proposition and gives the lower bound of $k$). 
The inequality $\eta < b_l$ yields 
  $ \eta + \sum_{k=j_1+1}^{l-1} b_k < b_l + \sum_{k=j_1+1}^{l-1} b_k = K - b_{l+1} < K$. 
\end{itemize}
Both cases imply that the value $\eta$ does not affect the condition in (\ref{eq:WleSecond}).  
Note that in either of both cases $d_1 = {\tt 0}$ and ${\tt 1}$,
  if we change the lower bound $j_1+1$ of the summation into $j_1$, 
  the summation become larger than $K$ irrespective of $\eta$
  because of $b_{j_1} = 2^{i+2}-1$.
We thus find (\ref{eq:WleSecond}) is constant for $\eta$ such that $0 \leq \eta < b_l$. 
For the third term of the right-hand side of (\ref{eq:Wle})
\begin{equation}
  \# \left\{ m \left| \ 0\leq l < m \leq r, \sum_{k=l}^{m-1} b_k \leq K + \eta \right. \right\}, 
  \label{eq:WleThird}
\end{equation}
$\sum_{k=l}^{j_2} b_k = K$ follows from proposition \ref{prop:ToRight} with $j_1= l - 1$, 
  where the upper bound $j_2$ is determined by $l$ in the proposition and gives the upper bound of $k$. 
In this case, $b_{j_2+1} = 1$ and $b_{j_2+2} = b_{j_2+3} = 2^{i} - 1$ hold because
  the consecutive elements $b_{j_2+1}$, $b_{j_2+2}$, and $b_{j_2+3}$ correspond to
  the strings $\beta d_1{\tt 2}$, $\beta d_1{\tt 1}$, and $\beta d_1{\tt 0}$ respectively.
Therefore we may evaluate (\ref{eq:WleThird}) as
\begin{equation*}
  \ds 
  \# \left\{ m \left| \ 0\leq l < m \leq r, \sum_{k=l}^{m-1} b_k \leq K \right. \right\} + \theta (\eta -1) + \theta (\eta-2^{i})
\end{equation*}
  for $0 \leq \eta < b_l = 2^{i+1}-1$ since the following properties hold;
\begin{itemize}
\item $\sum_{k=l}^{j_2+1} b_k = K+1 \leq K + \eta < \sum_{k=l}^{j_2+2} b_k $ \quad for $1 \leq \eta \leq 2^{i}-1$, 
\item $\sum_{k=l}^{j_2+2} b_k = K+2^{i} \leq K + \eta < \sum_{k=l}^{j_2+3} b_k $ \quad for $2^{i} \leq \eta < b_l = 2^{i+1}-1$. 
\end{itemize}

All cases for $\eta \neq 0$ in this subsection are summarized as follows:
\begin{eqnarray*}
  W(l, \eta) & = & W(l, 0)  + \theta (\eta -1) + \theta (\eta-2^{i} ) \\
  & \equiv & 2 + \theta (\eta -1) + \theta (\eta-2^{i} ) \bmod 3 \\
  & \not\equiv & 2 \bmod 3, 
\end{eqnarray*}
  where $0 \leq l \leq 2^{h+1}-(h+3)$ and $1 \leq \eta < b_l$. 
Because the sequence $\ds \left\{ b_l \right\}_{l=0}^{r-1}$ is palindromic,
  we obtain $M(0, W(\xi)-1) = 0$ for all integers $\xi \not\in J$.
\qed

\end{document}